\documentclass[12pt]{article}

\usepackage{bm}% bold math
\usepackage{amssymb}
\usepackage{enumerate}

\usepackage{graphicx} 

\begin{document}

%\documentclass[<options>]{elsarticle}

%\title{Experimental check of the model with light vector boson interacting with $L_{\mu} - L_{\tau}$ current at NA64}                    
%\title{The perspectives on  the search for light dark matter at NA64}               

%\title{The perspectives of light dark matter search   at NA64 experiment} 
\title{Combined search for light dark matter  with electron and muon beams at NA64}

\author{ S.N.~Gninenko$^{1}$, D.V.Kirpichnikov$^{1}$,  M.M.Kirsanov$^{1}$ and  N.V.~Krasnikov$^{1,2}$   
\\
$^{1}$ INR RAS, 117312 Moscow 
\\
$^{2}$ Joint Institute for Nuclear Research,141980 Dubna}

%\preprint{APS/123-QED}

%\title{Manuscript Title:\\with Forced Linebreak}% Force line breaks with \\

%\title{ Invisible  $K_L$ decays in the extensions of the $\nu$MSM}
%and constraints on new physics}
%\title{The  MiniBooNE anomaly and heavy neutrino decay}

\date{\today}% It is always \today, today,
%\date{June 17, 2009}% It is always \today, today,

\maketitle

\begin{abstract}
We discuss prospects  of searching for a dark photon ($A'$) which serves as mediator  
between Standard model (SM) particles and light dark matter (LDM) by using the combined 
results from  the NA64 experiment at the CERN SPS  running  in high-energy  electron (NA64e)  and muon (NA64$\mu$) modes.
We discuss  the most natural values and upper bounds on  the $A'$ coupling constant to LDM and show they are lying  in the range accessible at NA64. 
 While for the projected $ 5\times10^{12}$  electrons on target (EOT)   NA64e is able to probe the scalar and Majorana LDM scenarios,  the  combined NA64e and NA64$\mu$ results  with $\simeq 10^{13}$ EOT and  a few $10^{13}$ MOT, respectively,  will  allow covering significant region in the parameter space of the most interesting  LDM models.  This makes NA64e and NA64$\mu$  extremely complementary to each other and  increases significantly the discovery  potential of sub-GeV DM. 
 \end{abstract}

%\pacs{14.80.-j, 12.60.-i, 13.20.Cz, 13.35.Hb}% PACS, the Physics and Astronomy
%keywords:    {light dark matter, muon anomaly, light vector and scalar messengers}%Use showkeys class option if keyword
%\maketitle

\newpage
\section{Introduction}
Nowadays the most promising evidence in favour of a new physics beyond the SM is the observation of
Dark Matter (DM). In particular, various  DM models, for a review see e.g. \cite{Universe0} - \cite{Kolb},  which  motivate  
the existence of  light DM messengers with a mass $m_{\chi} \leq O(1)$~GeV  are of a great interest \cite{lightdark, jr}. 
The main idea  is that  in addition to gravity a new  interaction between  visible and dark sector can be
mediated by a new sub-GeV  vector or scalar  boson, as a review of the current and projected limits of  
LDM and other New Physics,  see e.g. Refs.~\cite{jr}-\cite{pbc}. 
%\par Anther possible hint in favour of new physics is the   muon $g_{\mu}-2$ anomaly, which is  
%Other possible hint in favour of new physics beyond the SM is 
%muon $g_{\mu}-2$ anomaly, namely the precise measurement of the anomalous magnetic
%moment of the positive muon from the
%Brookhaven AGS experiment \cite{g-2} gives a result which
%s $3.6 \sigma$ higher than the Standard Model (SM) prediction\footnote{Here $a_{\mu} \equiv \frac{g_{\mu} -2}{2}$.} 
%\begin{equation}
%a_{\mu}^{exp} - a_{\mu}^{SM} = (288 \pm 80) \times 10^{-11} \,.
%\end{equation} 
\par Among several  renormalizable LDM extensions of the SM, the   model with dark photon, where dark sector includes 
an abelian gauge field $A'_{\mu}$ (dark photon) is the most popular now. In these dark photon models,  dark sector 
interacts with the SM particles only through nonzero kinetic mixing  of the ordinary photon and dark photon,  
$-\frac{\epsilon}{2}F'_{\mu\nu}F^{\mu\nu}$.  In renormalizable models the DM particles interacting with the $A'$  
have spin $0$ or  $1/2$. Spin $1/2$ DM particles can be Majorana or pseudo-Dirac particles \cite{jr, Rev2018}. 
The annihilation cross-section for scalar or Majorana DM 
has $p$-wave suppression that allows to escape  the CMB bound 
\cite{p-wave,Planck} while for Dirac fermions the 
annihilation cross-section is $s$-wave that contradicts  to the CMB bound \cite{p-wave, Planck, Planck1}.   
For the model with pseudo-Dirac fermions \cite{Pseudo} it is also possible to avoid the CMB bound. 

Let us consider, as an example, charged scalar dark matter interacting with dark photons.
The charged dark matter field  $\chi$ interaction with   the $A'$ dark photon field is 
\begin{equation}
L_{\chi A'} = (\partial^{\mu}\chi - ie_DA'^{\mu}\chi)^{*}(\partial_{\mu}\chi - 
ie_DA'_{\mu}\chi) - m^2_{\chi}\chi^{*}\chi -  \lambda_{\chi}  (\chi^{*} \chi )^2  \,.
\label{lagrangian1} 
\end{equation} 
The nonrelativistic DM annihilation cross-section
  $\chi \bar{\chi} \rightarrow e^-e^+$  
has the form\footnote{Here we consider the case $m_{A'} > 2 m_{\chi}$, $m_{A'} \gg m_e$.}
\begin{equation}
\sigma_{an} v_{rel} = \frac{8\pi}{3} \frac{\epsilon^2\alpha\alpha_D m^2_{\chi}v^2_{rel}}
{(m^2_{A'} - 4 m^2_{\chi})^2 } \,.
\label{crsec}
\end{equation} 
Here $\alpha_D = \frac{e^2_D}{4\pi}$ is an analogue of the fine-structure constant $\alpha = 1/137$ 
for the DM particles interacting with DM photon.
We shall use a standard assumption  that in the hot early Universe DM is in equilibrium with 
ordinary matter \cite{Kolb}. During the Universe expansion the temperature decreases and at some 
temperature  the thermal decoupling of the DM occurs. Namely, at 
   freeze-out temperature $T_d$ the  cross-section of the annihilation $DM~particles\rightarrow ~SM~ particles$ 
 becomes too small to obey the equilibrium of the DM particles with 
the SM particles and the DM decouples. The experimental data are in favour of scenario          
with cold relic at which the freeze-out temperature $T_d$ is much lower 
than the mass of the DM particle. In other words, the DM particles decouple in 
the non-relativistic regime. The value of the DM annihilation 
cross-section at the decoupling temperature  determines the value of 
today's DM density in the Universe. In particular, relatively large annihilation cross-section 
leads to a low DM density. On the other hand, small annihilation 
cross-section leads to  DM overproduction. The observed 
value  of  DM density fraction, $\rho_{DM}/\rho_{c} \approx 0.23$, (here $\rho_{c}$ is a total  energy density of the Universe)  allows 
to estimate the  DM annihilation cross-section 
into the SM particles and hence to estimate the discovery potential of light DM for both  direct underground and accelerator experiments. 
One can roughly estimate the typical DM annihilation cross-section as \cite{Universe1}
%        \cite{Crude}\footnote{$1pbn = 2.57\times10^{-9}~GeV^{-2}$} the value of picobarn
\begin{equation}
<\sigma_{an}v_{rel}> = O(1)~\mbox{pb}.
\end{equation}
As a consequence of the formulae (2,3)  we can estimate the product 
$\epsilon^2\alpha_ D$ for fixed values $m_{A'}$ and $m_{\chi}$.
Note that fixed target NA64 experiment \cite{na64prd} 
uses the reaction
%\begin{equation}
%eZ \rightarrow eZA',  ~A' \rightarrow invisible \,
%\end{equation}
of the dark photon electroproduction on nuclei that allows obtaining only upper bounds on $\epsilon^2$ vs $m_{A'}$. Therefore,  to test the prediction 
for the $\epsilon^2\alpha_D$ we have to know either the $\alpha_D$ value or  at least its upper bound $\alpha_D \leq \alpha_o$. 
The arguments based on the use of the renormalization group and the assumption 
of the absence of the Landau pole singularity up to some scale $\Lambda$ allow 
to obtain upper limit on the coupling constant $\alpha_D$ \cite{Marciano}. The bound on $\alpha_D$ depends on the 
scale $\Lambda$ logarithmically. Moreover, the scale $\Lambda$ has to be 
larger than $1$~TeV \cite{Marciano}. So for fixed values of $m_{A'}$ and $m_{\chi}$ the knowledge of 
the upper bound on  $\alpha_D$ along with the requirement that the dark 
photon  model correctly reproduces the observed DM density  allows obtaining lower bound on $\epsilon^2$ as a function of $m_{A'}$ or $m_{\chi}$.  

In this paper we  discuss prospects of searching for $A'$ dark photon mediator  of  LDM  production 
at the NA64 experiment at CERN SPS  by using $\simeq$ 100 GeV   electron ( NA64e)  and muon ( NA64$\mu$)  beams.
 The rest of the paper is organized as follows. 
In Sec. 2 we discuss  upper bounds on $\alpha_D$ obtained from the requirement of 
the absence of Landau pole singularity for the effective coupling constant $\bar{\alpha}_D(\mu)$ up to some scale $\Lambda$. 
In Sec. 3  we estimate the  NA64e discovery potential of LDM and show that  with $\simeq 5\times10^{12}$ electrons on target (EOT)
 the experiment is  able to probe  the most natural parameter space of scalar and Majorana  LDM models.  
In Sec. 4  we estimate the  NA64$\mu$ discovery potential of LDM. 
 We  show that  NA64$\mu$ has better sensitivity to the $\gamma-A'$ kinetic mixing for the $A'$ masses $m_{A'} \gtrsim 100$ MeV 
  in comparison with NA64e, and that the  combined NA64e and NA64$\mu$  results obtained  with $\simeq 10^{13}$ EOT and  a few  $10^{13}$ MOT, respectively,  will  allow covering significant range of  natural 
parameter space of the LDM models including pseudo-Dirac LDM.  This 
makes the two approaches extremely complementary to each other and  increases 
significantly the discovery potential of NA64. 
%we investigate the NA64 discovery potential for  the dark photon model with light  dark matter.  
Sec. 5 contains  concluding remarks. In Appendix we collect the main formulae used for 
the  DM density calculations.  
%%%%%%%%%%%%%%%%%%%%%%%%%%%  
\section{Upper bound and range of  $\alpha_D$ }
%%%%%%%%%%%%%%%%%%%%%%%%%%%%%
One can  obtain upper bound on $\alpha_D$ by the requirement of  the absence of Landau 
pole singularity for the effective coupling constant 
$\bar{\alpha}_D(\mu)$ up to some scale  $\Lambda $ \cite{Marciano}.
One loop $\beta$ function for $\bar{\alpha}_D(\mu)$ is
\begin{equation}
\beta(\bar{\alpha}_D) = \frac{\bar{\alpha}_D^2}{2\pi}[\frac{4}{3}(Q^2_Fn_F +Q^2_S\frac{n_S}{4}) ] \,.
\end{equation} 
Here $\beta(\bar{\alpha}_D) \equiv \mu\frac{d\bar{\alpha}_D}{d\mu}$ and $n_F$ ($n_s$) is the number of fermions
 (scalars) with the $U^{'}(1)$ charge $ Q_F(Q_S)$. For the model with pseudo-Dirac fermion we introduce an additional scalar with $Q_S = 2$ 
to realize the splitting between fermion masses,  so 
 one loop $\beta$ function is $ \beta(\bar{\alpha}_D) =\frac{4\bar{\alpha}_D^2}{3\pi}$.
For the model with Majorana fermions,  we also introduce an additional scalar field with the charge $Q_S = 2$ and additional 
Majorana field to cancel $\gamma_5$ anomalies, so the $\beta$ function coincides with 
the  $\beta$ function for the model with pseudo-Dirac fermions. 
For the model with charged scalar matter, in order to create  nonzero dark photon mass, we have to introduce additional scalar field  with 
$Q_S = 1$,  so  one loop $\beta$ function is $\beta =  \alpha^2/3\pi$. 
From  the requirement that $\Lambda \geq 1$~TeV \cite{Marciano},  
we find that  $\alpha_D \leq 0.2$
for pseudo-Dirac and Majorana fermions and  $\alpha_D \leq 0.8$
for charged scalars \footnote{For smaller values of 
$\Lambda$ we shall have some charged particles with masses $\leq 1$~TeV that contradicts to the LHC bounds.}.
Here $\alpha_D $ is an effective low energy coupling at scale $\mu \sim m_{A'}$, i.e.
 $\alpha_D = \bar{\alpha}_D(m_{A'})$. In our calculations we used the value $m_{A'} = 10$~MeV.  
In the assumption  that dark photon model is valid up to Planck 
scale, i.e.   $\Lambda = M_{PL} = 1.2 \times 10^{19}$~GeV, we   find 
that for pseudo-Dirac and Majorana fermions  $\alpha_D \leq 0.05$ 
while for scalars $\alpha_D \leq 0.2$.
In  the SM the $SU_c(3)$, $SU_L(2)$ and $U(1)$ gauge coupling constants are equal to $\sim (1/30 - 1/50) $ at the Planck scale. 
One can show  that the   gauge coupling $\bar{\alpha}_D(\mu = M_{PL})$ is of the order of $\sim (1/30 - 1/50)$.  As a result, we find that the values in the range 
 $ \sim (0.014 - 0.02) $ are  the  most natural  for the low energy coupling constant $\alpha_D$.
%In the next section,   we shall use the values $\alpha_D =$ 0.1, 0.05 and 0.02 for numerical estimates.
 
 The expression (2) for the annihilation cross-section  is proportional to 
factor $K = (\frac{m^2_{A'}}{m_{\chi}^2} -4)^{-2}$ and in the resonance 
region $m_{A'} \approx 2m_{\chi}$  the 
 DM density bound on $\epsilon^2$   is proportional to $K^{-1}$. So  for $m_{A'} \approx 2m_{\chi}$ the bound on $\epsilon^2$ 
 becomes very weak \cite{Feng}.
% that makes the search for LDM  
%current and future accelerator bounds. 
It should be mentioned that  
in general  the values of $m_{A'}$ and $m_{\chi}$ are  arbitrary, so the case
$m_{A'} = 2m_{\chi}$   could be considered  as some fine-tuning. It is 
natural to require the absence of significant fine-tuning. 
Namely, we  require   that  $|\frac{m_{A'}}{2m_{\chi}} -1| \geq 0.25$, i.e. $m_{A'} \geq 2.5 m_{\chi}$.
In our estimates 
we  use  two values   $\frac{m_{A'}}{m_{\chi}} = 2.5   $ and   $\frac{m_{A'}}{m_{\chi}} = 3 $.
 We studied  DM models  with charged scalar, Majorana fermion and  pseudo-Dirac fermion \cite{Rev2018}. 
For the model with pseudo-Dirac DM 
we considered  the most difficult case  of small mass splitting $|\delta| \ll 1$\footnote{For a pseudo-Dirac fermion $\chi =(\eta, \phi)$ 
with $\eta$ and $\phi$ Weyl fermions the mass terms have the form \cite{Rev2018} 
 $L_m = -m_{\chi} \eta\phi - \frac{\Delta}{2}(\eta\eta + \phi\phi) + h.c.$. The mass eigenstates are $\chi_1 = \frac{i}{\sqrt{2}}(\eta - \phi)$ , $\chi_2 = \frac{1}{\sqrt{2}}(\eta + \phi)$ 
with masses $m_{1,2} = m_{\chi} \mp \Delta$. For $\delta \equiv \frac{\Delta}{m_{\chi}} \ll 1$  
DM density calculations 
coincide with the corresponding calculations for Dirac fermion DM.}.
Our calculations are based on the approximate formulae (10 - 13) presented in the Appendix. The results of our calculations for pseudo-Dirac DM density 
coincide with the 20~\% accuracy with the corresponding calculations of Ref.\cite{Rev2018}.  
%Besides we considered the models
% with charged scalar, Majorana fermion DM  and the model with pseudo-Dirac fermion DM. We considered the model with pseudo-Dirac fermion 
%DM for the most difficult case  of small mass splitting $|\delta| \ll 1$. 
%The limit $\delta \rightarrow 0$   corresponds to the case of Dirac fermion.

%%%%%%%%%%%%%%%%%%%%%%%%%%
\section{Projected LDM  sensitivity of NA64e }
\begin{figure}[tbh!!]
\begin{center}
\includegraphics[width=0.45\textwidth,height=0.37\textwidth]{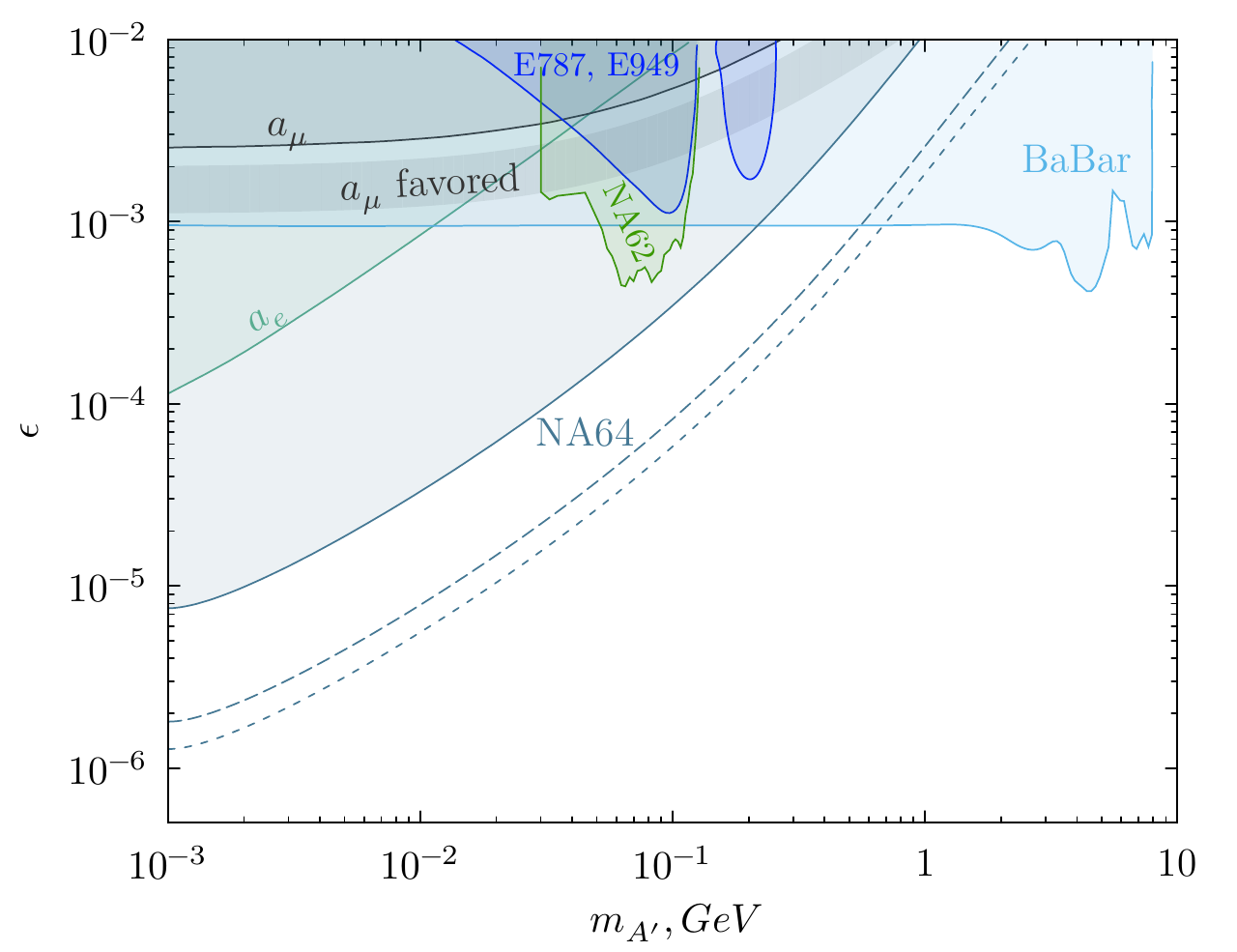}
\includegraphics[width=0.45\textwidth,height=0.4\textwidth]{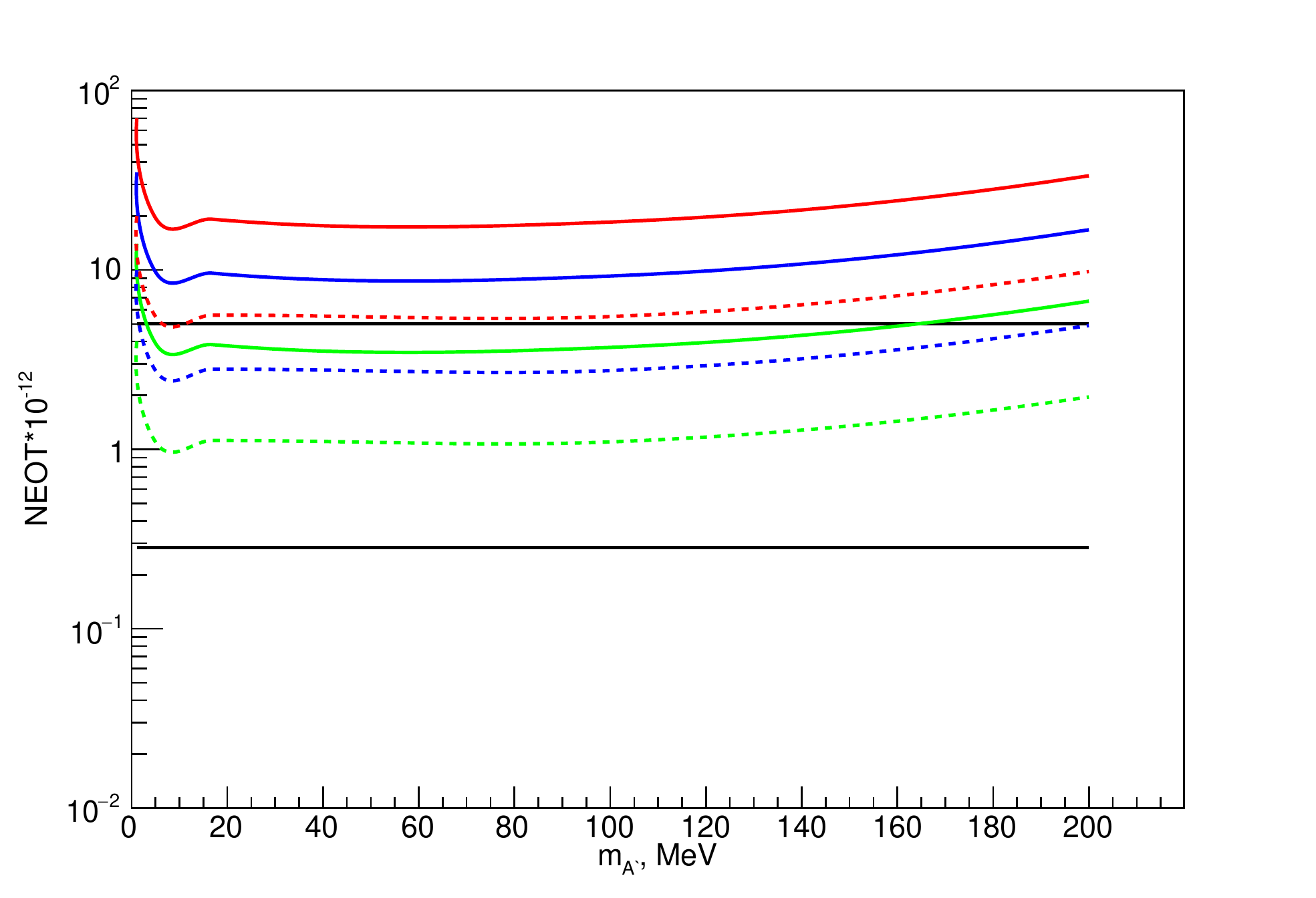}
\includegraphics[width=0.45\textwidth,height=0.4\textwidth]{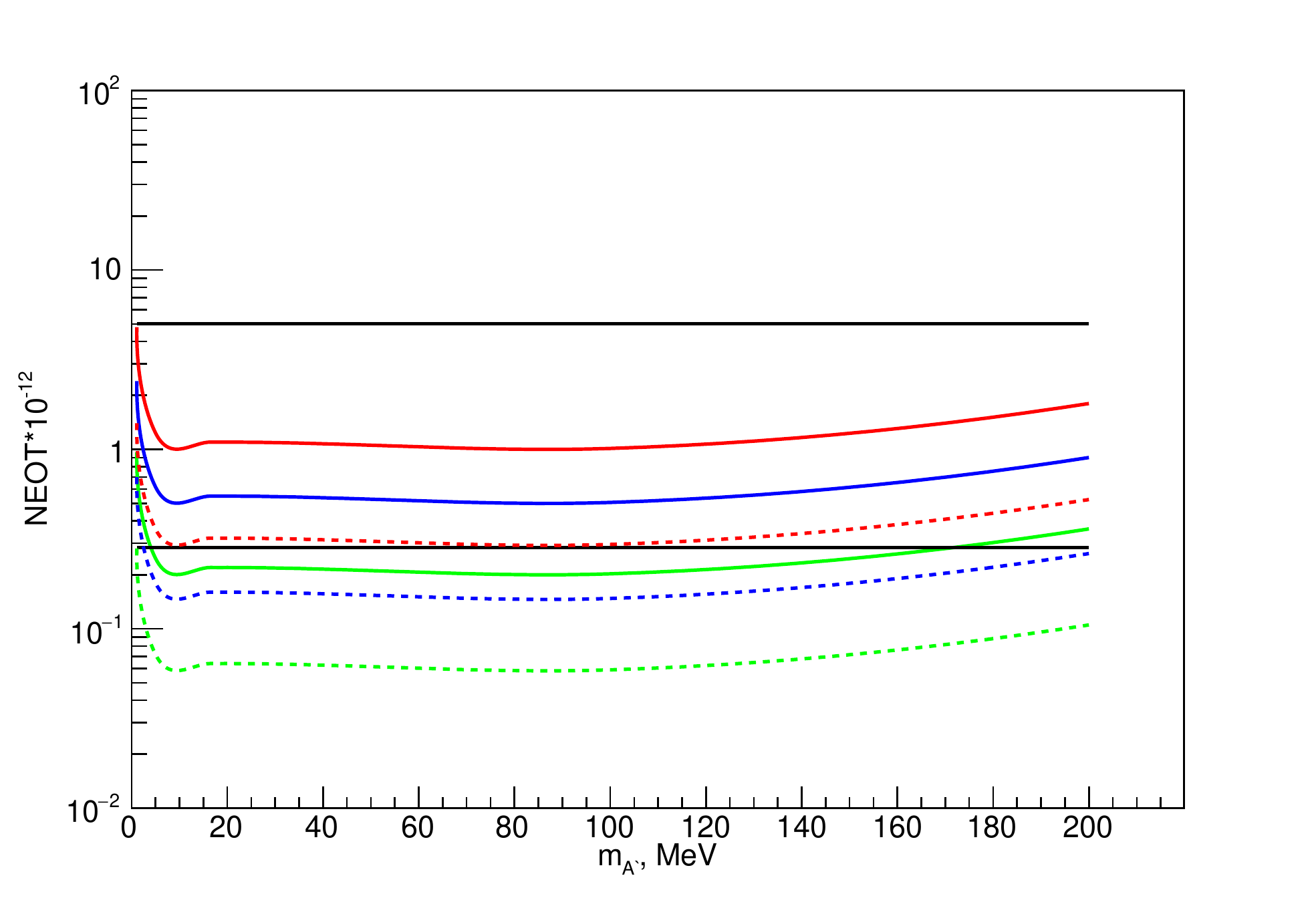}
\includegraphics[width=0.45\textwidth,height=0.4\textwidth]{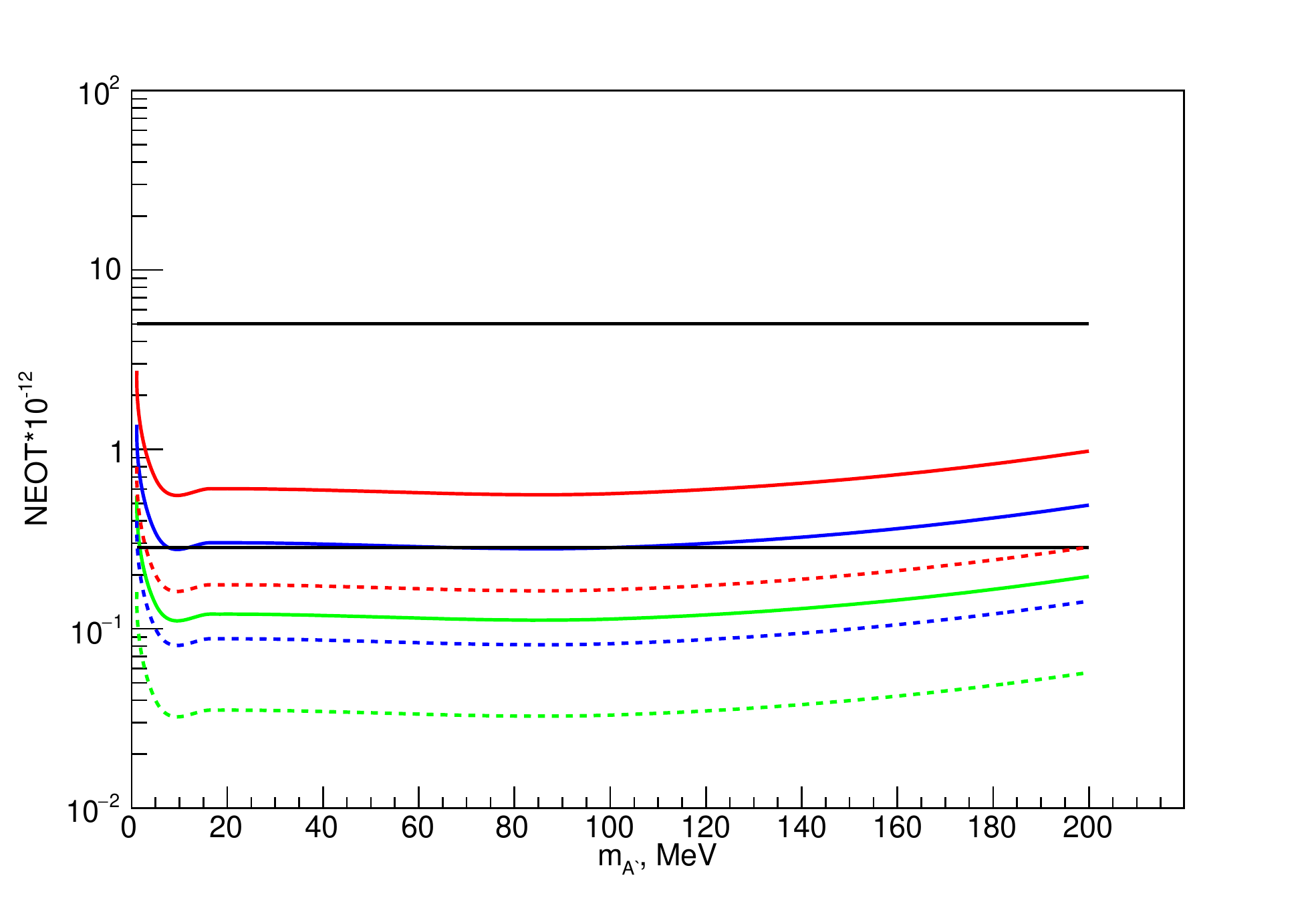}
\caption {The upper l.h.s. panel  shows NA64 90\% C.L. current (solid)  \cite{na64prl19}   bounds 
in the ($m_{A'},  \epsilon^2$) for $n_{EOT}=2.84 \times 10^{11}$; the projected sensitivities for $n_{EOT}=5\times 10^{12}$ 
and $n_{EOT}=10^{13}$ are shown by dashed and short-dashed lines 
respectively. 
The rest of the plots show the required number of EOT for the 90\% C.L. 
exclusion of the $A'$ with a given mass 
$m_{A'}$ in the ($m_{A'}, n_{EOT}\times10^{-12}$ )  plane for pseudo-Dirac  with 
$\delta \ll1$(the upper r.h.s. panel), Majorana (the lower l.h.s. panel), 
and scalar (the lower r.h.s. panel)   DM models for 
$\frac{m_{A'}}{m_{\chi}} = 2.5$ (solid),  and $= 3$ (dashed), and 
 $\alpha_{D} = $ 0.1 (red), 0.05 (blue),  
and  
0.02   (green). Upper(lower)  black lines correspond to $n_{EOT} = 5 \times 10^{12}(2.84 \times 10^{11})$. 
The curves under lower black line are excluded by last NA64 results  \cite{na64prd}.
\label{fig:excl-eps}}
\end{center}
\end{figure}

%%%%%%%%%%%%%%%%%%%%%%%%%
The NA64e experiment is designed for  a sensitive search for the $A'$ mediator of sub-GeV dark matter particle ($\chi$) production in the 
missing energy events from the reaction of 100 GeV electron scattering on heavy nuclei: 
\begin{equation}
e^- + Z \to e^- +Z + A'; A'\to \chi \chi  
\label{eq:react-e}
\end{equation}
at the CERN SPS \cite{Gninenko:2013rka,Andreas:2013lya}. 
After the long shutdown (LS2) stop at CERN the experiment plan to accumulate  
 $\gtrsim 5 \times 10^{12}$ EOT. The NA64e limits on mixing strength $\epsilon$ obtained from the 2016-2018 run with  $2.84 \times 10^{11}$ EOT \cite{na64prl19} and expected after the LS2 period assuming the zero-background case  \cite{pbc-bsm} are shown in the upper l.h.s. panel in 
 Fig.\ref{fig:excl-eps}.

The rest of the plots show the required number of EOT for the 
90\% C.L. exclusion of the $A'$ with a given mass 
$m_{A'}$ in the ($m_{A'}, n_{EOT}\times10^{-12}$ )  plane for pseudo-Dirac  with 
$\delta \ll1$ (the upper r.h.s. panel), Majorana (the lower l.h.s. panel), 
and Scalar
(the lower r.h.s. panel)   dark matter models 
for $\frac{m_{A'}}{m_{\chi}} = 2.5$ (solid),  and = 3 (dashed), and 
 $\alpha_{D} = $ 0.1 (red), 0.05  (blue),  
and 0.02 (green).
    As one can see,  NA64e is able to 
exclude  the most interesting and natural  LDM scenarios in the $A'$ mass range  
$1~MeV \leq m_{A'} \leq 150$~MeV
except the most difficult case of  pseudo-Dirac DM with 
$\alpha_D = 0.1$ and $\alpha_D = 0.05$, $\frac{m_{A'}}{m_{\chi}} = 2.5 $. 

%There are several different scenarios \cite{Rev2018} of  the dark photon model which 
%are  based 
%on $U(1)_{B-L}$  or  $U(1)_{B - 3e}$  gauge symmetries. As in the dark photon model, 
%the observed value of the DM density allows estimating the 
%coupling constant $\epsilon$ of new light $Z^`$ boson with an electron. The value of the $\epsilon$ parameter for such models 
%coincides with the $\epsilon$ value for dark photon model up to some factor $k \leq 3$ \cite{Rev2018}, 
%so NA64e can also test these scenarios. For instance, 
%for the model with $(B - L)$ vector interaction
% NA64e is able to exclude  scalar and Majorana dark matter scenarios in a way
% analogous to the case of dark photon model.    
%

%%%%%%%%%%%%%%%%%%%%%%%%%%%%%%%%%
\section{NA64$\mu$ projections for the $\gamma -A'$  mixing strength}
%%%%%%%%%%%%%%%%%%%%%%%%%%%%%%%%%%%%%
 The  NA64$\mu$  experiment \cite{GKM, prop-na64mu} is proposed to search for dark sector particles  weakly coupled to the muon, which could 
 explain the muon (g-2)$_\mu$ anomaly \cite{muonanomaly, PDG}. One of the good examples of such a particle, is  a new light vector  $Z'$ boson \cite{vecmuon1} - \cite{vecmuon8}, which  interacts predominantly with the  $L_{\mu} - L_{\tau}$ current\footnote{One loop corrections lead to 
nonzero interactions with electron, and other quarks and leptons \cite{GK2018}.}. Furthermore, 
  the $Z'$ could also  serve as a new leptophilic mediator of dark force between SM sector and dark matter, which is charged with respect to $U(1)_{L_{\mu}  - L_{\tau}}$. Moreover, this boson can be associated with
   the mechanism of the DM relic abundance~\cite{GK2018, krnjaic1, krnjaic2}.  
Another interesting possibility  involves muon-specific scalar mediator which could connect the visible and dark sectors and also 
account for the (g-2)$_\mu$ anomaly \cite{krasnikov2017, chen-na64mu-1, chen-na64mu-2}.   
\par   The  NA64$\mu$  plans  to perform a sensitive search for $L_{\mu}  - L_{\tau}$   $Z'$ as a  mediator  of sub-GeV dark matter particle ($\chi$) production in missing energy events from the reaction of 100-160 GeV muon scattering on heavy nuclei:
 \begin{equation}
\mu^- + Z \to  \mu^- + Z + Z'; Z' \to \nu\nu,  \chi\chi 
\label{eq:react-mu}
\end{equation}
at the CERN SPS \cite{GKM, prop-na64mu}. 

\par In the $A'$ dark photon model the interaction of dark photon with the leptons and quarks is given by $L_{A'} = e\epsilon A'_{\mu}J^{\mu}_{SM}$. 
Here,  $J^{\mu}_{SM} $ is  the electromagnetic current.  So, we see that  muons and electrons interact with the dark photon universally,
 with the same coupling constant.  Hence, similar to the reaction of Eq.(\ref{eq:react-e}), the dark photons will be also produced in the
  reaction of  
 Eq.(\ref{eq:react-mu}) with the same experimental signature of the missing energy.  
\begin{figure}[tbh!!]
\begin{center}
\includegraphics[width=0.5\textwidth,height=0.5\textwidth]{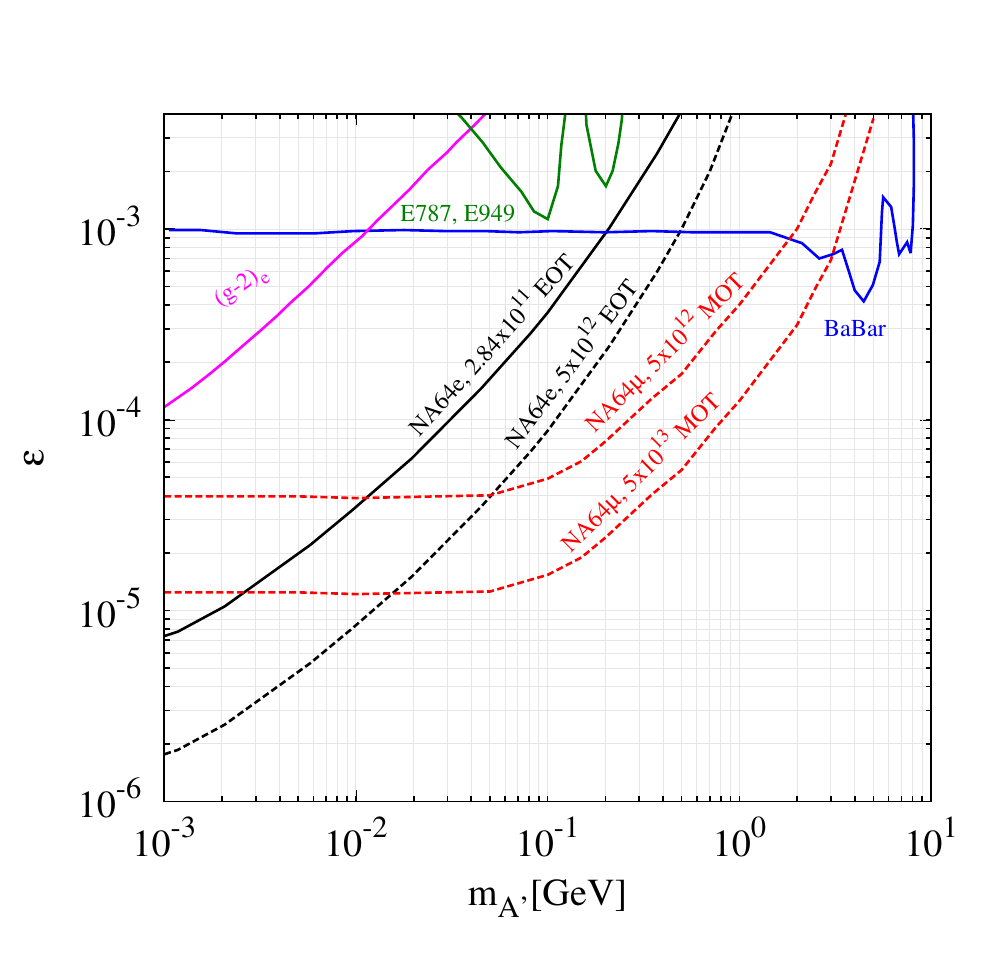}
\caption {The NA64e 90\% C.L. current \cite{na64prl19} and expected exclusion bounds  obtained with 
$2.84\times10^{11}$ EOT and  $5\times 10^{12}$ EOT, respectively,  in the ($m_{A'}, \epsilon$) 
plane.  The NA64$\mu$ projected  bounds calculated  for  $n_{MOT} = 5 \times 10^{12}$  and 
  $5\times 10^{13}$  are also shown.
 \label{fig:excl-muon}} 
\end{center}
\end{figure} 
% (up to logarithms)
% to $min (m^2_e, m^2_{A^`})^{-1}$ and for $m^2_{A^`} \gg m^2_e$ it is 
%proportional to $m^{-2}_{A^`}$. 
%The cross section for the dark photon muonproduction $\mu Z \rightarrow 
%\mu ZA^`$ is proportional (up to logarithms)
% to $min (m^2_{\mu}, m^2_{A^`})^{-1}$ and for $m^2_{A^`} \leq m^2_{\mu}$ 
% it is proportional to $m^{-2}_{\mu}$. 
%Besides the muon radiation length for muon is proportional to 
%$m^{-2}_{\mu} $ and it is bigger than the electron radiation length
 %by factor $(\frac{m_{\mu}}{m_e})^2$, so we can use more thick target for 
 %muon beam.
 For the $A'$ mass region  $m_{A'} \gg m_e$,  the total cross-section of the dark photon electroproduction $e Z \rightarrow e ZA'$  scales as    
$
\sigma^{e}_{A'}\sim \epsilon_{e}^2 / m_{A'}^2 
$. 
On the other hand,  for the dark photon masses, $m_{A'} \lesssim m_\mu$, the similar $\mu Z \rightarrow \mu ZA'$ cross-section can be approximated  in the bremsstrahlung-like limit as
$
\sigma^{\mu}_{A'}\sim \epsilon_{\mu}^2 /  m_\mu^2
$.
Let us now  compare expected sensitivities of the $A'$ searches with NA64e and NA64$\mu$  experiments for the
same number $\simeq 5\times 10^{12}$   particles on target.  Assuming the same signal efficiency 
 the number of $A'$ produced by the 100 GeV electron and muon beam 
 can  approximated, respectively,   as follows
\begin{equation}
N^{e}_{A'} \approx   \frac{\rho N_{av}}{A} \cdot n_{EOT} L^{e} \sigma_{A'}^{e}, \qquad 
N^{\mu}_{A'} \approx   \frac{\rho N_{av}}{A} \cdot n_{MOT} L^{\mu} \sigma_{A'}^{\mu}, 
\label{NmuNe}
\end{equation}
where  $L^{e} \simeq X_0$  and $L^{\mu}\simeq 40 X_0$ are  the 
typical distances that are  passed by an electron  and muon, respectively,   before producing
the $A'$ with the energy $E_{A'} \gtrsim 50$ GeV   in the NA64 active Pb target of the total thickness of $\simeq 40$ radiation length ($X_0$) 
\cite{GKM, prop-na64mu}.  
The detailed comparison of the calculated $A'$ sensitivities of NA64e and NA64$\mu$ 
 is shown  in Fig.\ref{fig:excl-muon},  where the 90\% C.L. limits on the mixing $\epsilon$ are shown for 
a different number of particles on target for both the NA64e and NA64$\mu$ experiments. 
%  $5\times 10^{12}$ and    $5\times 10^{13}$ MOT together with the limits from NA64e. 
The limits were obtained for the background free case by using exact-tree-level 
(ETL) cross-sections rather than the Weizsacker-Williams (WW) ones calculated  for NA64e  in Ref.\cite{ETL}, 
 and for  the NA64$\mu$ case in this work. The later are shown in  Fig.~\ref{fig:etlCSmuon}  as a function of $E_{A'}/E_\mu$ for the Pb target and mixing value $\epsilon=1$.  One can see that in a wide range of masses, $20 \mbox{ MeV} \lesssim m_{A'} \lesssim 1 \mbox{ GeV}$, the  
total WW cross-sections  are larger by a factor $\simeq 2$ compared to the ETL ones. 
 As the result, the typical  limits on $\epsilon$ 
 for the ETL case are worse by about  a factor $\simeq 1.4$  compared to the WW case. 
 \begin{figure}[tbh!!]
\begin{center}
\includegraphics[width=0.75\textwidth]{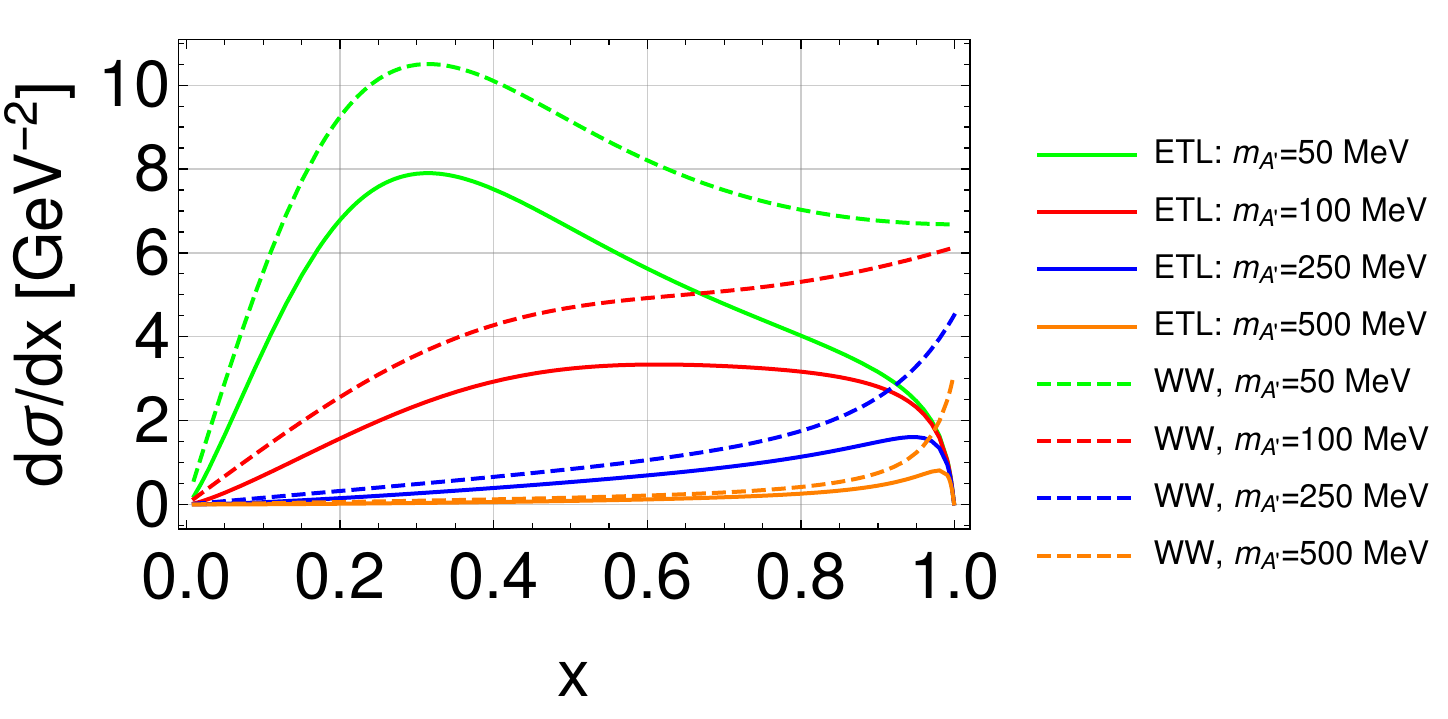}
\caption {Cross-section of dark photon 
production by muons as a function of $x=E_{A'}/E_\mu$ for various masses $m_{A'}$ and $\epsilon =1$. Solid lines represent ETL cross-sections and
dashed lines show the cross-sections calculated in WW approach.
\label{fig:etlCSmuon}} 
\end{center}
\end{figure}  
For  $n_{EOT} = n_{MOT}=5 \cdot 10^{12}$ the sensitivity of NA64e 
  is enhanced for  the mass range $m_e \ll  m_{A'} \simeq 100$ MeV 
%Therefore, from~(\ref{CSeCompSmallM}), (\ref{CSmuCompSmallM}) and (\ref{NmuNe}) one has $\epsilon_{(\mu)} \gtrsim \epsilon_{(e)}$ for $m_{A'} \lesssim 30$~MeV as shown in Fig.~\ref{fig:excl-muon}.
while for the $A'$ masses  $ m_{A'} \gtrsim 100$ MeV NA64$\mu$  allows 
obtaining a more stringent limits  on $\epsilon^2$  
compared to  NA64e.

%The bound on $\epsilon^2$ were calculated by using the results of Refs. \cite{GKM, prop-na64mu}
%in the assumption that the average muon beam energy is $E_{\mu} = 160~GeV$,  the $Pb$ target length is $L = 40~X_0$,  and the background free case.    As one can see,  
%for $m_{A'} \geq 20-30~MeV$ the use of muon beam  with   $\gtrsim 10^{13}$ accumulated MOT  will allow to obtain more strong bound on $\epsilon^2$  than the use of electron beam with $ 5 \times 10^{12}$ EOT.
%Here we would like to emphasize the increased sensitivity of NA64$\mu$  allowing us to obtain more stringent limits  on the mixing strength  
%$\epsilon^2$  for the dark photon masses $ m_ {A '} \geq$ 20-30 MeV than in NA64e. 

\section{Combined LDM sensitivity of NA64e and NA64$\mu$ }
\begin{figure}[tbh!!]
\begin{center}
\hspace{-0.5cm}{\includegraphics[width=.45\textwidth,height=0.4\textwidth]{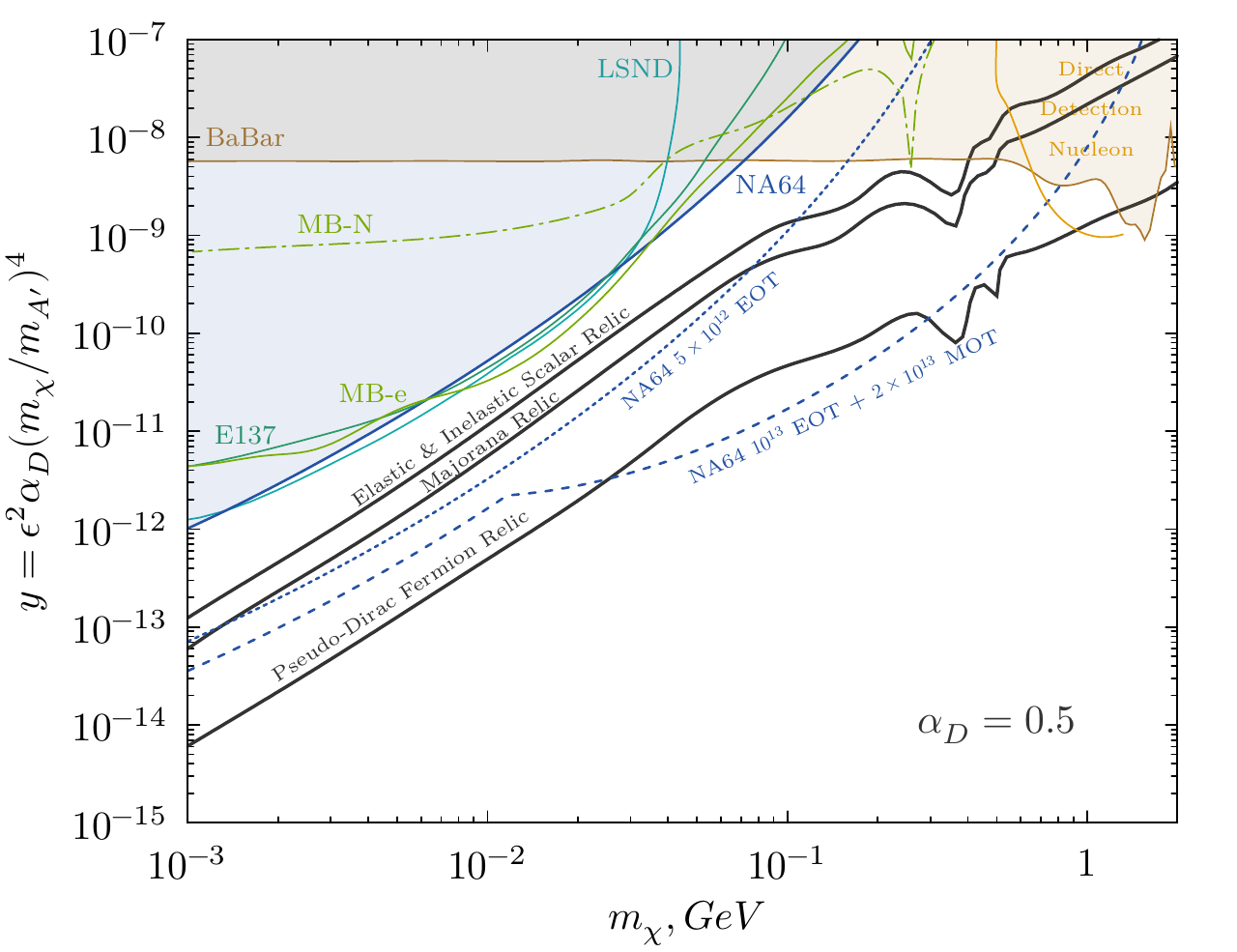}
\includegraphics[width=.45\textwidth,height=0.4\textwidth]{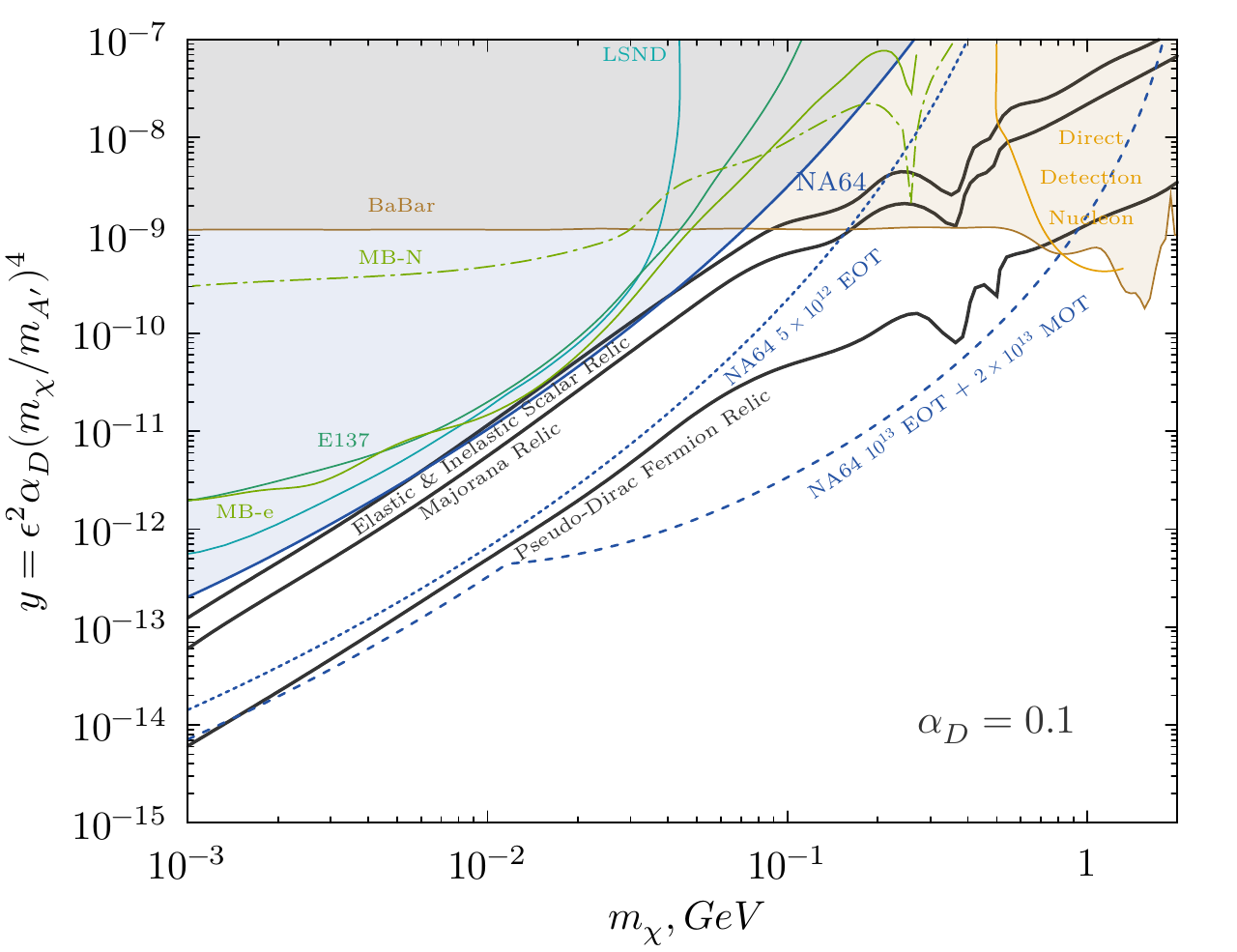}}
\includegraphics[width=.48\textwidth,height=0.4\textwidth]{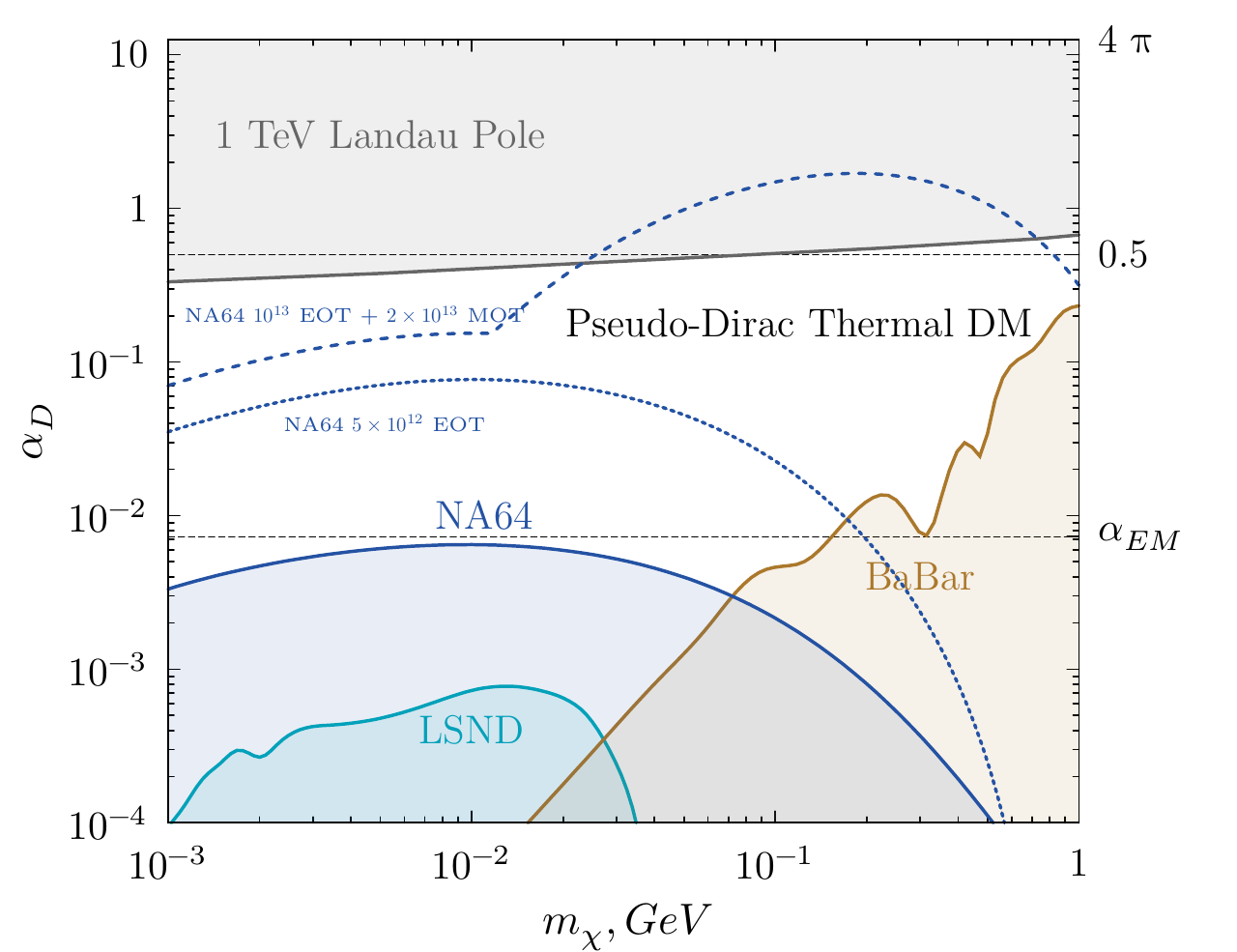}
\includegraphics[width=.48\textwidth,height=0.4\textwidth]{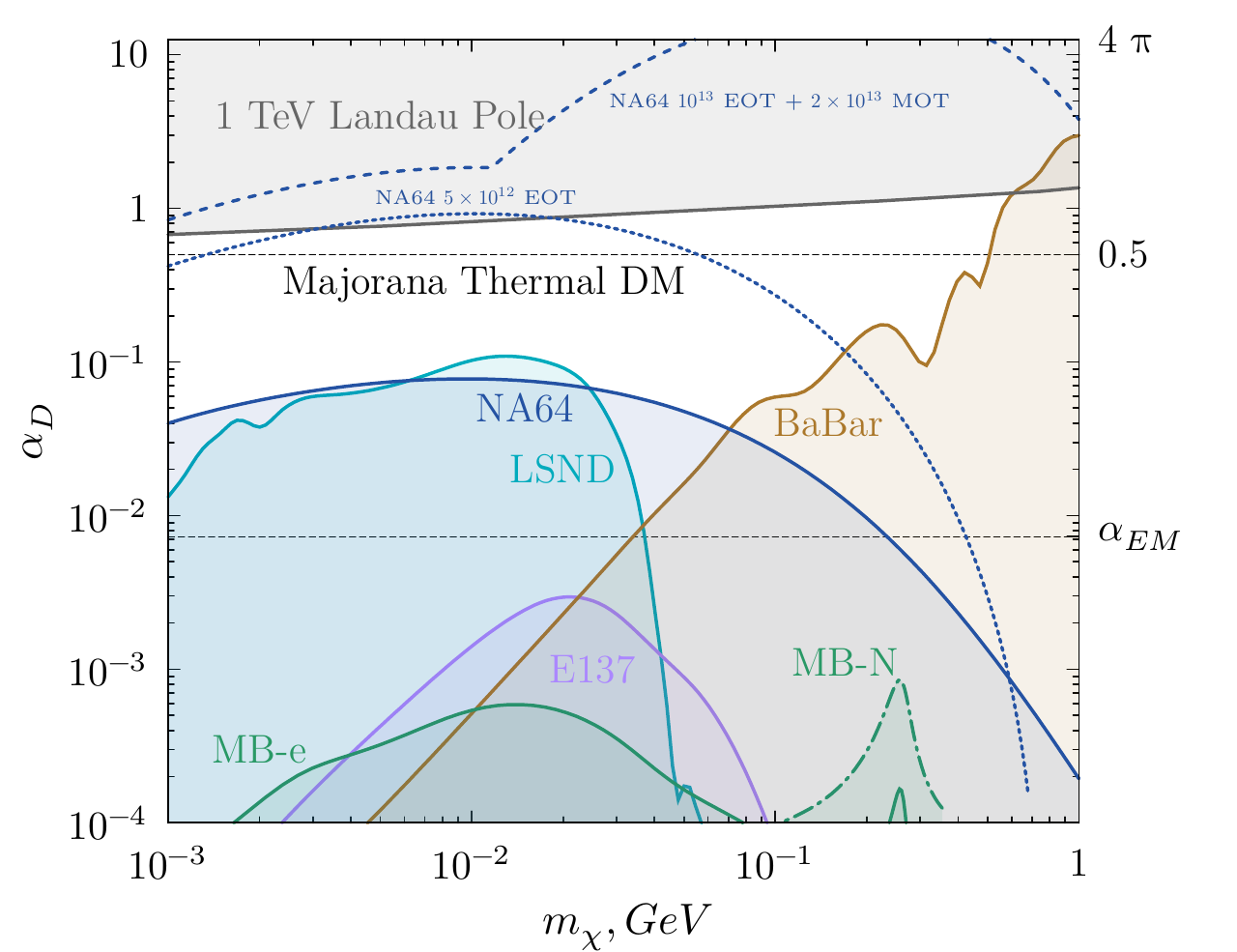}
\caption {The  NA64 90\% C.L. current (solid) \cite{na64prl19} and  expected (dotted light blue)  exclusion bounds  for 
 $ 5\times10^{12}$ EOT  in the ($m_{\chi}, y$) and ($m_{\chi}, \alpha_D$)  planes. The combined limits from NA64e and NA64$\mu$ are also shown 
 for  $10^{13}$ EOT plus $2\times 10^{13}$ MOT (dashed  blue).
 The black solid curves show the favoured parameters to account for the observed  DM relic density for the scalar, pseudo-Dirac and Majorana type of light thermal DM, see e.g. Ref.~\cite{Rev2018}.
  The limits are  calculated for $\alpha_D=0.1$ and  0.5,  and   $m_{A'}=3m_{\chi}$.
The results are also  shown in comparison with bounds obtained from  the results of the 
LSND~\cite{report2, deNiverville:2011it, Batell:2009di},   E137 \cite{e137th}, BaBar \cite{babarg-2}  and MiniBooNE \cite{ minib2018} experiments.
  \label{fig:comb-limit}} 
\end{center}
\end{figure}

The                 estimated  NA64e and NA64$\mu$  limits on the $\gamma-A'$ mixing strength, allow us to set the 
combined NA64e and NA64$\mu$ constraints  on the LDM models, which are shown 
in the $(y;~m_\chi$)  plane in Fig.\ref{fig:comb-limit}.
 As discussed in Sec. I, as a result of  
the $\gamma - A'$ mixing  the cross-section of the DM particles annihilation into the SM particles is proportional to $\epsilon^2$. Hence using  
constraints on the  DM annihilation  cross-section 
%freeze out
% (resulting in  Eq.(\ref{crsec}), and obtained limits on mixing 
%strength of Figs.~\ref{fig:excl-eps}, ~\ref{fig:excl-muon} 
one  can derive constraints 
 in the ($y \equiv \epsilon^2 \alpha_D (m_\chi/m_{A'})^4; ~m_{\chi}$) plane and restrict the LDM models with the masses  $m_{\chi} \lesssim 1$~GeV. 
%Here, the variable   $y$ is  $y=\epsilon^2 \alpha_D (m_\chi/m_{A'})^4$. 
\par The combined limits obtained  from the 
data sample of the 2016 \cite{na64prl,na64prd}, and 2017, 2018 runs \cite{na64prl19} and expected from the run after the LS2   are shown in  the top panels of 
Fig.~\ref{fig:comb-limit} together with combined limits from NA64e and NA64$\mu$ for  
 $10^{13}$ EOT and $2\times 10^{13}$ MOT, respectively.
%The favoured parameter curves for scalar,  pseudo-Dirac (with a small splitting)  and Majorana  scenario  of  light DM   
%taking into account  the observed relic DM density, see e.g. \cite{alex, pbc-bsm}. 
%The limits were  calculated  using  Eq.(8) from Ref.\cite{na64prd}
%   under the  assumption $\alpha_D= 0.1, 0.005$, and  $m_{A'}=3 m_{\chi}$, here $m_\chi $ stands for the LDM 
%particle's masses,  either scalars or fermions. 
The plots show also the comparison of our results with the limits of other experiments.  
%The choice of $\alpha_D = 0.1$ is  based on arguments for 
%the running of the dark gauge coupling, presented in Sec. I.  
It  should be 
 noted  that   the $\chi$-yield in the NA64 case scales as $\epsilon^2$ rather than  $\epsilon^4 \alpha_D$ as in beam dump experiments. Therefore, 
for sufficiently small values of $\alpha_D$  the NA64 limits  will be much stronger. This is 
illustrated in the upper right panel of  Fig.~\ref{fig:comb-limit},  where the NA64 limits are shown for  
 $\alpha_D = 0.1$. One can see,  that  for this, or smaller, values of   $\alpha_D$,  the direct search for LDM  at NA64e with $5\times10^{12}$ EOT
  excludes  the scalar and Majorana  models of  the LDM production via vector mediator with $\frac{m_{A^{'}}}{m_{\chi}} = 3$
   for the full    mass region  up to $m_\chi \lesssim 0.2$ GeV.  
While being combined with the NA64$\mu$  limit, the NA64 will  exclude the models with $\alpha_D \leq 0.1$ 
for the  entire mass region up to $m_\chi \lesssim 1$ GeV. 
 So we see that for the full mass range $m_{\chi} \lesssim 1$ GeV 
 the obtained combined NA64e and NA64$\mu$ bounds are more stringent than the limits obtained from the results of  NA64e 
 that allows probing the full sub-GeV DM parameter space.

\section{Conclusions}
%%%%%%%%%%%%%
In this paper we considered the NA64 discovery perspectives   of  sub-GeV  thermal dark matter by running the experiment in electron and muon modes at the CERN SPS.
Remarkably,  that with the statistics accumulated during years 2016-2018 NA64 already starts probing  the sub-GeV DM parameter space for the conventional value of 
$\alpha_D=0.1$ \cite{na64prl19}. While  with $ 5\times10^{12}$  EOT   NA64e is able to test the scalar and Majorana  LDM scenarios for $\frac{m_{A'}}{m_{\chi}} \geq 2.5$,  
the  combined NA64e and NA64$\mu$ results  with $\gtrsim 10^{13}$ EOT and  $2\times 10^{13}$ MOT, respectively,  will  allow to fully explore   the parameter space of 
other  interesting LDM models like pseudo-Dirac DM model or the model with new light vector boson $Z'_{B-L}$.
% due to the better NA64$\mu$ sensitivity to the $\gamma-A'$ kinetic mixing in the  $m_{A'} \gtrsim 30$ MeV  $A'$ mass range. 
 This makes NA64e and NA64$\mu$  extremely complementary to each other, as well as 
 to the planned LDMX experiment \cite{ldmx}, and greatly increases the  NA64  discovery potential of  sub-GeV DM.  
 
There are several different scenarios \cite{Rev2018} of  the dark photon model which 
are  based 
on $U(1)_{B-L}$  or  $U(1)_{B - 3e}$  gauge symmetries. As in the dark photon model, 
the observed value of the DM density allows estimating the 
coupling constant $\epsilon$ of new light $Z'$ boson with an electron. The value of the $\epsilon$ parameter for such models 
coincides with the $\epsilon$ value for dark photon model up to some factor $k \leq 3$ \cite{Rev2018}, 
so NA64e can also test these scenarios. For instance, 
for the model with $(B - L)$ vector interaction
 NA64e is able to exclude  scalar and Majorana dark matter scenarios in a way
 analogous to the case of dark photon.

% There are several alternatives \cite{Rev2018} to  the dark photon model based 
%on the use of gauge symmetries like $U(1)_{B-L}$  or  $U(1)_{B - 3e}$. As in the dark photon model 
%the observed value of the DM density allows to estimate the 
%coupling constant $\epsilon$ of new light $Z^`$ boson with electron. The value of the $\epsilon$ parameter for such models 
%coincides with the $\epsilon$ value for dark photon model up to some factor $k \leq 3$ \cite{Rev2018}, 
%so NA64e can also test such  models. For instance, 
%for the model with $(B - L)$ vector interaction
% NA64e is able to exclude  scalar and Majorana dark matter scenarios in full analogy with the case of dark photon model.    

However it should be stressed that for $m_{A'} \approx 2m_{\chi}$ the DM annihilation cross-section (2) is proportional to
$(m^2_{A'} - 4 m^2_{DM})^{-2}$. As a consequence the predicted value of the $\epsilon^2$ parameter is proportional to 
$(\frac{m^2_{A'}}{4m^2_{\chi}} -4)^2 $ that can reduce the predicted $\epsilon^2$ value by 2 - 4 orders of magnitude in comparison with the 
often used value  $\frac{m_{A'}}{m_{\chi}} =3$ \cite{Feng}. It means  that  NA64 experiment and other future experiments like LDMX \cite{ldmx} 
are  not able to test the region
 $m_{A'} \approx 2m_{\chi}$ completely\footnote{The values of $m_{A'}$ and $m_{\chi}$ are  arbitrary, so the case
$m_{A'} \approx  2m_{\chi}$   could be considered  as some fine-tuning. It is 
natural to assume  the absence of significant fine-tuning. 
In this paper we  required   that  $ m_{A'} \geq 2.5 m_{\chi}$.}.

%\section{Acknowledgements}
{\bf Acknowledgements}\\
% We thank organizers of the PBC Workshops at CERN: C. Valle, M. Lamont, J. Jaeckel,  members of the PBC BSM and other working groups, in particular J. Bernhard, M. Brugger, L. Gatignon, G. Lanfranchi, J. Jaeckel,  A. Rozanov  for their support, discussions  and  valuable comments. 
 We are indebted to Prof. V.A. Matveev and  our colleagues from  the NA64  Collaborations  for many for useful  discussions. We thank N.~Toro for 
 valuable comments. 
  We  also would like to thank R. Dusaev 
 for his help in preparing of Fig.4. 

\appendix{\bf{Appendix. Basic formulae for DM density}}

%Namely, the annihilation cross-section leading to the correct DM density is 
%estimated to be 
The relic density of DM in the standard scenario is  obtained 
by solving the Boltzmann equation 
\begin{equation}
\frac{dn_{d}}{dt} + 3H(T)n_{d} = - <\sigma v_{rel}>(n^2_{d} - n^2_{d,eq})\,.
\end{equation}
Here 
\begin{equation}
n_{d}(T) = \int \frac{d^3p}{2\pi^3} f_{d}(p,T) \,
\end{equation}
and $f_{d}(p,T)$ is the dark matter distribution function.

The dark matter relic density can be numerically estimated as \cite{Kolb}
\begin{equation}
\Omega_{DM}h^2 = 0.1\Bigl(\frac{(n+1)x_f^{n+1}}{(g_{*s}/g^{1/2}_*)}\Bigr)\frac{0.876\cdot 10^{-9}\mbox{GeV}^{-2}}{\sigma_0} \,,
\end{equation}
where $<\sigma v_{rel}> = \sigma_o x^{-n}_f$ 
and 
\begin{equation}
x_f = c - (n + \frac{1}{2}){\rm ln}(c) \,,
\end{equation}
\begin{equation}
c = {\rm ln}(0.038(n+1)\frac{g}{\sqrt{g_*}}M_{Pl}m_{\chi}\sigma_0) \,.
\end{equation}
Here $x_f =\frac{m_{\chi}}{T_{d}}$,  $n = 0(1)$ for $s(p)$-wave annihilation and $g_*$, $g_{*s}$ are 
the effective relativistic energy and entropy degrees of freedom.  
If DM particles differ from DM antiparticles $\sigma_o = \frac{\sigma_{an}}{2}$.
The  requirement that the dark photon model reproduces correct value of the DM density allows to estimate  $\alpha_D$ 
as a function of $\epsilon $, $m_{A'}$ and $m_{\chi}$, namely \cite{Marciano}:
 \begin{equation}
\alpha_D \simeq 0.02 f(m_{A^{'}}, m_{\chi})\cdot  \Bigl( \frac{10^{-3}}{\epsilon}\Bigr)^2\Bigl( \frac{m_{A'}}{100~ \mbox{MeV}}\Bigr)^4
\Bigl( \frac{10~\mbox{MeV}}{m_\chi}\Bigr)^2   \,
\label{alphad}
\end{equation}
% which are shown in lower panels of Fig.~\ref{fig:comb-limit}  in the ($\alpha_D$;$m_{\chi}$) plane.
 For the  pseudo-Dirac DM 
with $\frac{m_{A'}}{m_{\chi}} = 3$ and $\delta \ll 1$ the estimates based on the use  of the  formulae (10 - 12)  
lead to $f = 0.25 - 0.4 $ 
  at   $1~\mbox{MeV} \leq m_{\chi} \leq 100$~MeV while analogous estimate for  Majorana DM gives $f = 3 - 5$. 

%We used conservative values $f = 0.25(3)$ for pseudo-Dirac(Majorana) DM.}   , see

%and small splitting ,   the limits  in the lower left panel of Fig.~\ref{fig:comb-limit} were calculated by taking 
 %the value $f=0.25 $.  

%One can see that for the full mass range $m_{\chi} \lesssim 0.05$ GeV 
% the obtained combined NA64e and NA64$\mu$ bounds are more stringent than the limits obtained from the results of  NA64e 
% allowing to probe almost the full parameter space.  

% By  taking  into account  that the annihilation cross sections for Majorana fermions 
%are approximately two times larger compared  to that for the scalar case,
% The limits for the Majorana case   shown in the lower right panel of Fig.~\ref{fig:comb-limit} were calculated by setting $f=3$
%\footnote{More accurate  calculations based on the use of semianalytic formulae (11 - 13) for DM density 
%lead to $f = 0.25 - 0.4(3-5)$ for 
%pseudo-Dirac(Majorana) DM   at   $1~MeV \leq m_{DM} \leq 100~MeV$. We used conservative values $f = 0.25(3)$ for pseudo-Dirac(Majorana) DM.}   , se%e \cite{na64prd}.
% Similar to pseudo-Dirac case the   combined NA64e and NA64$\mu$ limits exclude the full remained parameter area.
%    Note,  that  new constraints for the large pseudo-Dirac fermion splitting   
% can also be derived. They will be more stringent than for the case of the small splitting 
% and similar to the one obtained for the Majorana case. 

\end{document}